\documentclass[a4paper,british,12pt]{iopart}
\usepackage[T1]{fontenc}
\usepackage[latin9]{inputenc}
\usepackage{color}
\usepackage{amssymb}
\usepackage{graphicx}
\usepackage{esint}
\usepackage{ulem}
\usepackage{iopams}
\usepackage{setstack}
\usepackage{amstext}
\usepackage{bbold}
\usepackage{soul}
\usepackage{babel}
\usepackage{subfig}
\usepackage{babel}
\usepackage{epstopdf}
\usepackage[numbers,square,sort&compress]{natbib}
\usepackage[unicode=true,pdfusetitle,bookmarks=false,breaklinks=false,pdfborder={0 0 1},backref=false,colorlinks=false]{hyperref}

\PassOptionsToPackage{normalem}{ulem}
\makeatletter

\newcommand{\binom}[2]{{#1 \choose #2}}
\newcommand{\ket}[2][]{{|#2\rangle_{#1}}}

\newcommand{\bra}[2][]{{}_{#1}\langle #2|}

\newcommand{\proj}[2][]{\ket{#2}_{#1}\bra{#2}}

\definecolor{nblue}{rgb}{0.3,0.3,1.0}
\definecolor{ngreen}{rgb}{0.2,0.7,0.2}
\definecolor{nred}{rgb}{0.9,0.1,0}
\definecolor{norange}{rgb}{0.8,0.5,0}

\@ifundefined{showcaptionsetup}{}{ \PassOptionsToPackage{caption=false}{subfig}}
\makeatother

\begin{document}

\title{Optimal lossy quantum interferometry in phase space}
\author{Andrei B. Klimov$^{1}$, Marcin Zwierz$^{2}$, Sascha Wallentowitz$%
^{3} $, Marcin Jarzyna$^{4}$, and Konrad Banaszek$^{2,4}$}
\address{$^{1}$Departamento de F\'{i}sica, Universidad de Guadalajara, Revoluci\'{o}n
1500, 44410 Guadalajara, Jalisco, Mexico}

\address{$^{2}$Faculty of Physics, University of Warsaw, Pasteura 5, PL-02-093
Warszawa, Poland}

\address{$^{3}$Instituto de F\'{i}sica, Pontificia Universidad Cat\'{o}lica de Chile,
Casilla 306, Santiago 22, Chile}

\address{$^{4}$Centre of New Technologies, University of Warsaw, Banacha 2c, PL-02-097 Warszawa, Poland}

\ead{m.jarzyna@cent.uw.edu.pl}

\begin{abstract}
We analyse the phase space representation of the optimal measurement of a
phase shift in an interferometer with
equal photon loss in both its arms. In the local phase estimation scenario with a fixed number
of photons, we identify features of the spin Wigner function that warrant sub-shot noise precision,
and discuss their sensitivity to losses. We derive the asymptotic form of an integral
kernel describing the process of photon loss in the phase space in the limit of
large photon numbers. The analytic form of this kernel allows one to
assess the ultimate precision limit for a lossy interferometer.
We also provide a general lower bound on the quantum Fisher information in terms of spin Wigner functions.
\end{abstract}

\maketitle

\section{Introduction}

\label{sec::intro}

The purpose of optical interferometry is to measure a phase difference between two
light beams with a high precision. When coherent laser light is used, the precision of
such a phase measurement, quantified by
the standard deviation $\Delta\tilde\varphi$, is limited by the shot noise equal to $%
\Delta\tilde\varphi= 1/\sqrt{N}$, where $N$ is the mean photon number in
the input beam \cite{Ou1997, Giovannetti2004}. The shot-noise limit can be surpassed using
non-classical light characterized by squeezed quadrature fluctuations \cite{Caves1981, Yurke1986, Ono2010, Seshadreesan2011} or alternatively the
N00N state, which is a coherent superposition of a fixed number of
$N$ photons all in one or another arm of the interferometer \cite{Sanders1989, Boto2000}. The use
of a N00N state allows one to reach the ultimate bound on the precision derived only from the principles of quantum mechanics, known as the Heisenberg limit \cite{Lee2002, Giovannetti2004, Afek2010} and given by $\Delta\tilde\varphi= 1/N$, which
offers a qualitatively improved scaling with the photon number $N$ in comparison to the shot noise limit. However,
whereas this result holds for an idealised interferometric setup with unit
transmission, in any experimental implementation one needs to take into
account imperfections with the ubiquitous example of photon loss. In
realistic scenarios when attenuation cannot be neglected, N00N states are no longer
optimal with respect to precision and should be replaced by other multiphoton superposition states \cite{DornDemkPRL09,DemkDornPRA09}.
In certain regimes a near-optimal
precision can be attained by using squeezed states \cite{Ono2010, Seshadreesan2011, LIGO2011, LIGO2013, Demkowicz2013}. The non-trivial
structure of optimal states and their sensitivity to the loss rate are
rather curious and deserve a study to obtain intuition about their specific
features.

In the present paper we provide a phase-space description of photon losses in a two-arm interferometer. As the main tool we will use the spin Wigner function \cite{Stratonovich1956, Agarwal1981, Varilly1989, Dowling1994}, naturally defined for two-mode field states in sub-manifolds with a fixed photon number \cite{Demkowicz2015}. It will be shown that such an approach gives intuitive pictorial insights into the form of the optimal states for quantum-enhanced
interferometry in the presence of loss \cite{DornDemkPRL09,DemkDornPRA09}
allowing one to identify graphically features that are behind enhanced precision of phase estimation. 
We will analyze the situation when the loss strength is identical in both the arms of the interferemeter, which leads to an essential simplification of the algebraic structure of the problem. In particular, the loss operation commutes with any SU(2) transformation of the two-mode system, allowing an elegant
description of the photon-loss process as a convolution of the input Wigner function with an integral kernel.
Furthermore, we obtain an explicit asymptotic form of the
integral kernel when the input photon number is large,  $N \gg 1$, and 
the most likely numbers of photons both lost and those appearing at the
interferometer output are much greater than one. This asymptotics provides an
intuitive insight into the ultimate, optimized over all $N$-photon input states, precision limit for a lossy
interferometer.

Several classes of states relevant to quantum interferometry have been discussed using
the phase space picture in \cite{CombWiseJOB05}. Our focus here is to
include the loss transformation in the phase space description and to
identify structures that warrant sub-shot noise precision in the presence of
photon loss. The derived convolution map for the photon loss allows one also
to analyse how losses affect the sensitivity of the input state to the
measured phase shift. We also provide a quantitative link between the
precision of local phase estimation in which we estimate a small deviation from a known operating point and the Wigner formalism.
Specifically, we show that the quantum Fisher information which defines this precision
can be lower-bounded by phase-space integrals containing expressions quadratic in respective Wigner functions and their derivatives.

This paper is organized as follows. In Sec.~\ref{sec:Interferometery} we briefly review quantum interferometry with a fixed number of photons. The loss transformation is discussed in Sec.~\ref{Sec:LossTransformation}. In Sec.~\ref{sec:Wigner} we apply the Wigner phase space formalism to describe the optimal states for local phase estimation. Sec.~\ref{sec:PhaseSpaceLoss} analyses the effect of losses on the structure of those states. We derive an approximate analytical form of the loss transformation in the limit of large photon numbers which provides an intuitive argument for the asymptotic scaling of the phase estimation precision. Finally, Sec.~\ref{Sec:Conclusions} concludes the paper.

\noindent
\begin{figure}[tbp]
\centering \includegraphics[width=0.5\columnwidth]{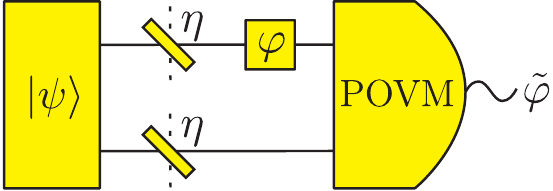}
\caption{A general interferometer, where an $N$-photon two-mode input state $|%
\protect\psi\rangle$ undergoes losses, denoted by fiticious beam splitters with transmission $\eta$, and a phase shift $\protect\varphi$ resulting in an output state
$\rho_\varphi$. The output state is subjected to a general quantum
measurement described mathematically by a positive-operator-valued measure (POVM).
Finally, based on the measurement outcomes the value of $\protect\varphi$
is estimated with an estimator function $\tilde{\protect\varphi}$.}
\label{fig::interferometer}
\end{figure}

\section{Interferometry with definite photon-number states}
\label{sec:Interferometery}

Consider the standard interferometer depicted in Figure \ref{fig::interferometer}, where a fixed number of $N$ photons distributed
between two bosonic modes, that shall be described by the annihilation
operators $\hat{a}$ and $\hat{b}$, is used to probe a relative phase shift $%
\varphi$ between these modes. A general pure $N$-photon two-mode input state
$\hat{\varrho} = \ket{\psi}\bra{\psi}$ for such an
interferometer can be written as a state vector in the angular momentum
representation
\begin{equation}
|\psi \rangle =\sum_{m=-J}^{J}c_{m}|J,m\rangle ,  \label{Eq:psi}
\end{equation}%
where the total spin is $J=N/2$ and individual basis states read in the
number-state representation
\begin{equation}
|J,m\rangle \equiv \frac{(\hat{a}^\dagger)^{J+m}(\hat{b}^\dagger)^{J-m}}{%
\sqrt{(J+m)!(J-m)!}}\ket{\text{vac}}= |(J+m)_a (J-m)_{b}\rangle ,
\label{amb}
\end{equation}%
i.e.\ the modes $a$ and $b$ carry $J+m$ and $J-m$ photons, respectively, and
$m=-J,-(J-1),\ldots ,0,\ldots ,J$. In the above expression, the ket $\ket{\text{vac}}$ denotes the vacuum state of the electromagnetic field.
The relative phase delay between the arms $a$ and $b$ of the interferometer
corresponds to the application of a unitary transform
\begin{equation}
\hat{U}_{\varphi }=\exp (-i \varphi \hat{n}_a) ,  \label{eq:Uphi}
\end{equation}%
where $\hat{n}_{a}=\hat{a}^{\dagger }\hat{a}$ is the photon number operator
in the sensing arm $a$. The light in the interferometer also experiences
losses described by a completely positive trace preserving map $%
\Lambda(\cdot)$, whose explicit form will be given in the next section.
This map commutes with the phase shift so we can assume without a loss of generality that losses occur prior to phase shift. Finally, the phase-shifted two-mode
output state $\hat\varrho_\varphi = \hat{U}_{\varphi} \Lambda(\hat{\varrho} )%
\hat{U}^\dagger_{\varphi}$ is subject to a general quantum measurement
formalized as a positive-operator-valued measure (POVM). Based on the
outcomes of these measurements the value of $\varphi $ is estimated with an
estimator $\tilde{\varphi}$. In the case of local phase estimation, the
performance of the measurement can be quantified with the quantum Cram\'{e}r--Rao inequality \cite{Helstrom1976, Holevo1982, Braunstein1994, Giovannetti2011}, which provides a lower bound on the standard deviation
\begin{equation}
\Delta\tilde{\varphi} \geq\frac{1}{\sqrt{F_{Q}[\hat{\varrho}_{\varphi}]}},\quad F_Q\left[\hat{\varrho}_{\varphi}\right]=\Tr \bigl( \hat{\varrho}_{\varphi}\hat{L}_{\varphi}^2\bigr)
\label{eq:est-error}
\end{equation}
for any unbiased estimator $\tilde{\varphi}$ of the phase shift, optimized
over all possible POVMs, that is, over all conceivable quantum measurements
\cite{Helstrom1976,Holevo1982,Braunstein1994}. Here $F_{Q}[\hat{\varrho}%
_{\varphi}]$ is the quantum Fisher information, which in principle
depends on the true value of $\varphi$, and $\hat{L}_{\varphi}$ is an operator called symmetric logarithmic derivative, implicitly given by equation $\frac{d\hat{\varrho}_{\varphi}}{d\varphi}=\frac{1}{2}\{\hat{\varrho}_{\varphi},\hat{L}_{\varphi}\}$, where $\{\,.\,,\,.\,\}$ denotes an anticommutator. Note that Cram\'{e}r--Rao inequality in general is saturable only for asymptotically large datasets, that is in the limit of infinite number of repetitions of the experiment \cite{Braunstein1992}. Nevertheless, since for every finite number of repetitions it bounds the performance of any unbiased estimator from below, it is a useful tool for analysis of estimation protocols and we employ it in this work. Crucially, quantum Fisher information can be considered not only as a bound on precision but also as a measure of sensitivity of the state to a phase shift, thus relating these two concepts (see \ref{Sec:Superfidelity}).

In the lossless scenario when the attenuation map $\Lambda(\cdot)$ is
trivial and given by identity, the density operator of the output state
remains pure, $\hat{\varrho}_{\varphi}=|\psi_{\varphi}\rangle\langle\psi_{%
\varphi}|$ where $\ket{\psi_\varphi} = \hat{U}_\varphi\ket{\psi}$, and the
quantum Fisher information takes the well-known form \cite{Braunstein1996, Demkowicz2015},
\begin{equation}
F_{Q}[|\psi_{\varphi}\rangle\langle\psi_{\varphi}|]=4(\langle\dot{\psi}%
_{\varphi}|\dot{\psi}_{\varphi}\rangle-|\langle\psi_{\varphi}|\dot{\psi}%
_{\varphi}\rangle|^{2})\ \ \mbox{with}\ \ |\dot{\psi}_{\varphi}\rangle=\frac{%
d|\psi_{\varphi}\rangle}{d\varphi}.
\end{equation}
Using the explicit form of the unitary transformation $\hat{U}_\varphi$
given in Eq.~(\ref{eq:Uphi}), quantum Fisher information can be written as $%
F_{Q}[|\psi_{\varphi}\rangle\langle\psi_{\varphi}|] = 4 \bigl( \bra{\psi}%
\hat{n}_a^2 \ket{\psi} - (\bra{\psi}\hat{n}_a \ket{\psi})^2\bigr)$, i.e.\
quadrupled variance of the photon number in the sensing arm $a$. This
variance is maximized for a N00N state of the form \cite%
{Lee2002,Seshadreesan2011}
\begin{equation}
|\mbox{N00N}\rangle=\frac{1}{\sqrt{2}}(|J,J\rangle+|J,-J\rangle) \equiv
\frac{1}{\sqrt{2}}(\ket{N_a 0_b}+\ket{ 0_a N_b}),  \label{Eq:N00Ndef}
\end{equation}
for which one has $F_{Q}=N^{2}$ and consequently the estimation error is $%
\Delta\tilde{\varphi} = 1/N$. The scaling of this lower bound is known as
the Heisenberg limit providing a $\sqrt{N}$ enhancement in phase estimation
with respect to the shot-noise limit $\Delta\tilde{\varphi}=1/\sqrt{N}$ \cite%
{Demkowicz2015}.

\section{Loss transformation}

\label{Sec:LossTransformation}

Apart from the fact that N00N states are difficult to prepare experimentally
for higher $N$ \cite{Afek2010}, in the presence of photon loss they are no longer optimal
for local phase estimation \cite{Gilbert2008, DornDemkPRL09}. The output state in a lossy scenario becomes mixed
and it is necessary to use the general expression for the quantum Fisher information
involving the symmetric logarithmic derivative given in Eq.~(\ref{eq:est-error}).

Let us start by writing explicitly the photon loss map $\Lambda(\cdot)$ in
the angular momentum representation. We will assume that the power
transmission for both the interferometer arms is the same and
equal to $\eta$, i.e.\ the photon loss probability in each arm is $1-\eta$.
For an arbitrary two-mode bosonic state $\hat{\varrho}$, the map can be
written as a double sum with two indices: $L$ corresponding to the total
number of lost photons and $l$ specifying the number of photons lost from
the arm $a$:
\begin{equation}
\Lambda(\hat{\varrho})=\sum_{L=0}^{\infty} \sum_{l=0}^{L}\hat{K}_{l}^{(a)}%
\hat{K}_{L-l}^{(b)}\hat{\varrho}\hat{K}_{L-l}^{(b)\dagger }\hat{K}%
_{l}^{(a)\dagger }.  \label{KO1}
\end{equation}
Here $\hat{K}_{l}^{(a)}$ and $\hat{K}_{L-l}^{(b)}$ are Krauss operators
removing respectively$l$ photons from the mode $a$ and $L-l$ photons from
the mode $b$. They are given explicitly by \cite%
{DornDemkPRL09,DemkDornPRA09,WasiBanaPRA07}
\begin{eqnarray}
\hat{K}_{l}^{(a)} & = & \frac{1}{\sqrt{l!}}(1-\eta )^{l/2}\eta ^{\hat{a}%
^{\dagger }\hat{a}/2}\hat{a}^{l},  \nonumber \\
\hat{K}_{L-l}^{(b)} & = &\frac{1}{\sqrt{(L-l)!}}(1-\eta )^{(L-l)/2}\eta ^{%
\hat{b}^{\dagger }\hat{b}/2}\hat{b}^{L-l}.  \label{Eq:Krauss}
\end{eqnarray}

Suppose now that the input state $\hat\varrho$ contains exactly $N$ photons.
The overall probability $p_{L}^{N}$ that $L$ photons have been lost is given
by the binomial distribution
\begin{equation}
p_{L}^{N}={{{\binom{N }{L}}}}\eta ^{N-L}(1-\eta )^{L}.
\label{Eq:LambdaBinTP-1}
\end{equation}
The conditional transformation of the input state for the loss of $L$
photons in total can be described by a trace preserving completely positive
map $\Lambda^N_L(\cdot)$, which maps the spin-$(J=N/2)$ space onto a spin-$%
(J^{\prime }=N^{\prime }/2)$ space, where $N^{\prime }=N-L$ is the remaining
number of photons. Furthermore, subspaces with different $J^{\prime }$ are
completely incoherent as the knowledge of the remaining photon number is in
principle contained in the environment. Consequently, when starting from a
state with a definite photon number $N$, the state after losses can be
written as a direct sum
\begin{equation}
\Lambda(\hat{\varrho}) = \bigoplus_{L=0}^{N} p^N_L \Lambda^N_L(\hat{\varrho}%
).  \label{Eq:DirectSum}
\end{equation}
The explicit form of the maps $\Lambda^N_L(\hat{\varrho})$ can be obtained
by rearranging Eq.~(\ref{KO1}) using the explicit form of the Krauss
operators in Eq.~(\ref{Eq:Krauss}). After straightforward algebra one
arrives at
\begin{equation}
\Lambda _{L}^{N}(\hat{\varrho})=\frac{(N-L)!}{N!}\sum_{l=0}^{L}{{{\binom{L }{%
l}}}}\hat{a}^{l}\hat{b}^{L-l}\hat{\varrho}(\hat{a}^{l}\hat{b}%
^{L-l})^{\dagger}.  \label{Eq:LossTransformation}
\end{equation}%
where the action of the operator monomial $\hat{a}^{l}\hat{b}^{L-l}$ is
given in the angular momentum representation as
\begin{equation}
\hat{a}^{l}\hat{b}^{L-l}|J,m\rangle =\sqrt{\frac{(J+m)!(J-m)!}{%
(J+m-l)!(J-m-L+l)!}}\,|J^{\prime },m-l+\frac{L}{2}\rangle .
\label{Eq:lossinangularmomentumrep}
\end{equation}
It is worth noting that although the map $\Lambda _{L}^{N}(\cdot)$ changes
the spin from $J=N/2$ to $J^{\prime }=N^{\prime }/2=(N-L)/2$, it commutes
with the phase shift $\hat{U}_\varphi \cdot \hat{U}_\varphi^\dagger$ and
more generally any SU(2) transformation of the input state.
Further, the map $\Lambda _{L}^{N}(\cdot)$
does not depend explicitly on the transmission $\eta$, which enters only the the overall probabilities
$p_{L}^{N}$ defined in Eq.~(\ref{Eq:LambdaBinTP-1}).

Using the decomposition given in Eq.~(\ref{Eq:DirectSum}), the quantum
Fisher information for the output state $\hat\varrho_\varphi = \hat{U}%
_{\varphi} \Lambda(\hat{\varrho} )\hat{U}^\dagger_{\varphi}$ can be written
as a sum of contributions from individual spin subspaces,
\begin{equation}
F_Q[\hat{\varrho}_\varphi] = \sum_{L=0}^{N} p^N_L F_Q[\hat{U}_\varphi
\Lambda^N_L(\hat{\varrho}) \hat{U}_\varphi^\dagger ].  \label{Eq:FQ=sumL}
\end{equation}
This decomposition is possible owing to the fact that the probabilities $%
p^N_L$ do not depend on the phase shift $\varphi$. For a pure input state
containing exactly $N$ photons, $\hat{\varrho}=\proj{\psi}$ with $\ket{\psi}$
defined in Eq.~(\ref{Eq:psi}), all the terms on the right hand side of Eq.~(%
\ref{Eq:FQ=sumL}) depend on a single set of probability amplitudes $c_m$, $%
m=-J, \ldots, J$. This makes the task of optimizing Eq.~(\ref{Eq:FQ=sumL}) a
nontrivial matter. The most straightforward way to find the solution is to
resort to numerical means \cite{DornDemkPRL09, DemkDornPRA09, Jarzyna2013, Macieszczak2014, Macieszczak2014a, Jarzyna2015}.

\begin{figure}[tbp]
\includegraphics[width=1\columnwidth]{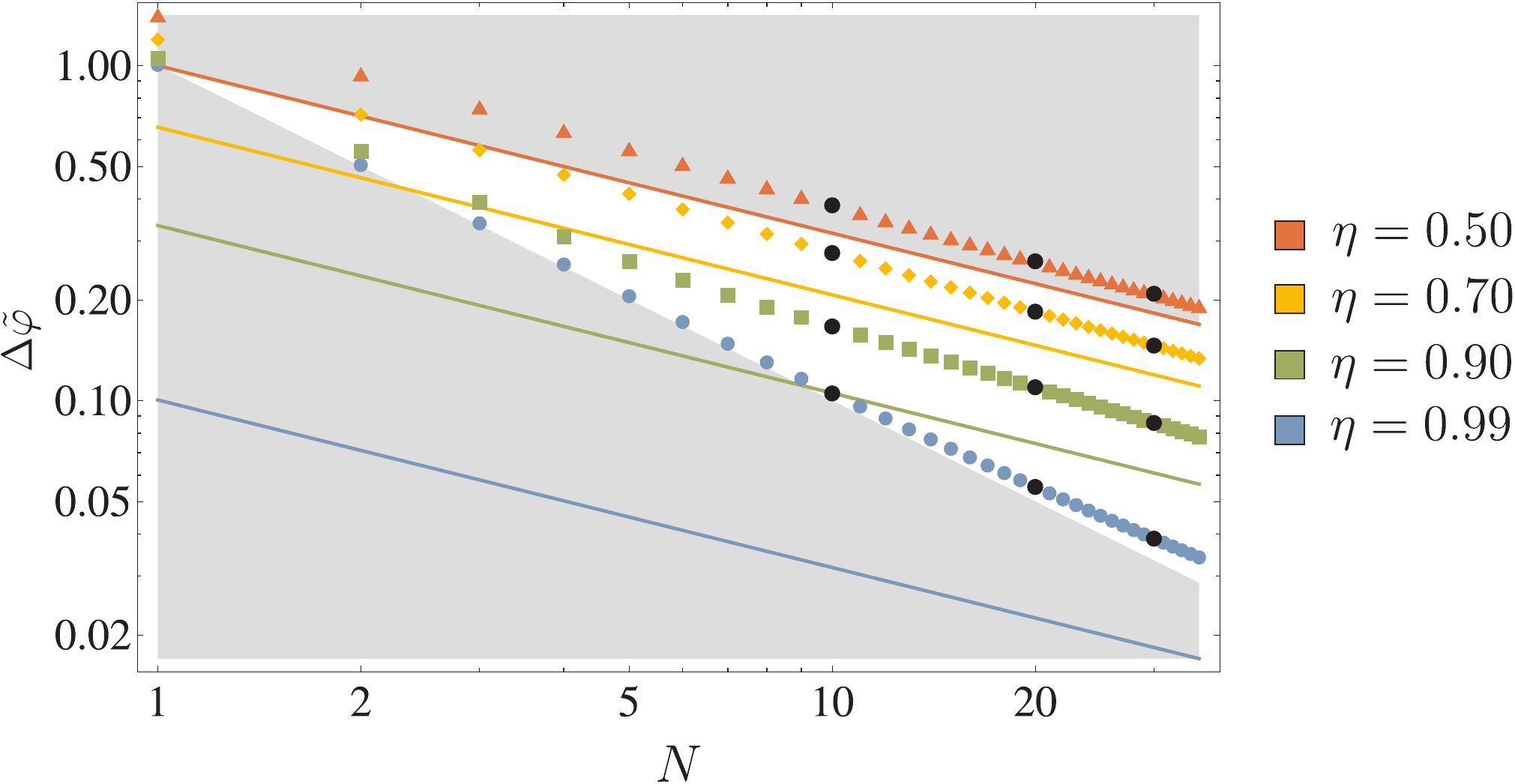}
\caption{The quantum Cram\'{e}r--Rao bounds on precision of phase estimation in the presence of a
two-arm photon loss optimized over input states plotted against the number of probe photons $N$ for the
four values of photon loss coefficient $\protect\eta$. The markers depict
the exact optimal precision bounds for each value of $\protect\eta$ with
solid lines representing the asymptotic precision bounds as per (\protect\ref%
{eq:asymptotic_loss}). The gray areas correspond to precision regions
which for $\eta=1$ represent worse than shot-noise scaling (the upper one) and better
than the Heisenberg scaling (the lower one). The black dots are the
specific values of photon number $N=10,\ 20\ \mathrm{and}\ 30$ for which we
 provide graphical representations of their angular Wigner functions $%
W_{\hat{\protect\varrho}_{\mathrm{opt}}}^{N}(\protect\theta,\protect\phi)$ in Fig.~(\ref{fig::optimalWigner}).}
\label{Fig::loglog}
\end{figure}

In Fig.~\ref{Fig::loglog} we depict using the log-log scale the ultimate
precision $1/\sqrt{F_Q[\hat{\varrho}_\varphi]}$ obtained by the numerical
optimization of Eq.~(\ref{Eq:FQ=sumL}) over coefficients of the input state Eq.~(\ref{Eq:psi}) as a function of the photon number $N$
for several values of the interferometer transmission $\eta=50\%$, $70\%$, $%
90\%$, and $99\%$. As a reference, we show also as edges of grey outer
regions the Heisenberg limit $1/N$ and the standard shot noise limit $1/%
\sqrt{N}$ for the lossless case when $\eta=1$. Note that the shot noise
limit for a lossy interferometer is given by $1/\sqrt{\eta N}$, which in the
logarithmic scale of Fig.~\ref{Fig::loglog} corresponds to shifting the shot
noise boundary by the amount corresponding to $1/\sqrt{\eta}$. All numerical calculations for this and following plots were done using Mathematica 9.

Fig.~\ref{Fig::loglog} shows that for low photon numbers the benefit of
using optimal states is quite substantial, whereas for higher $N$ the
precision seems to follow shot-noise-type scaling, although improved by a
multiplicative factor compared to the shot noise limit $1/\sqrt{\eta N}$.
This observation is confirmed by a rigorous asymptotic analysis of the
optimal precision scaling reported in \cite{Knysh2011, Escher2011,Demkowicz2012, Knysh2014}, which
shows that for sufficiently large $N$ the precision approaches
\begin{equation}
\Delta \tilde{\varphi} \approx \sqrt{\frac{1-\eta }{\eta N}}.
\label{eq:asymptotic_loss}
\end{equation}
These asymptotic expressions are shown as solid lines in Fig.~\ref%
{Fig::loglog}. It is seen that with decreasing transmission $\eta$ the
convergence to the asymptotic regime occurs for lower photon numbers $N$.

In the following we will analyze in detail the phase space picture of
optimal two-mode states for $N=10,20,30$ and four values of the transmission
parameter, marked as black dots in Fig.~\ref{Fig::loglog}. These examples
cover different regimes, ranging from approaching closely the asymptotic
limit given in Eq.~(\ref{eq:asymptotic_loss}) for $\eta=50\%$ to remaining
almost at the Heisenberg limit for $\eta=99\%$.

\section{Wigner function of optimal states}
\label{sec:Wigner}

Let us now discuss the Wigner phase space representation of optimal input states
for quantum enhanced interferometry in the presence of two-arm photon loss.
Our focus will be to identify graphically features that are behind the enhanced precision of phase estimation.

A general $N$-photon two-mode input probe state $\hat{\varrho}$ treated as a
spin system with total angular momentum $J=N/2$ can be elegantly represented
as a quasiprobability distribution function on a unit sphere $\mathcal{S}_{2}$ parameterized
with $\Omega=(\theta,\phi)\in\mathcal{S}_{2} $ using the spin Wigner
function $W_{\hat{\varrho}}^{N}(\theta,\phi)$ \cite%
{Stratonovich1956, Agarwal1981, Varilly1989}
\begin{equation}\label{eq:Wigner_general}
W_{\hat{\varrho}}^{N}(\Omega)=\Tr[\hat{w}_{N}(\Omega)\hat{\varrho}].
\end{equation}
The operator $\hat{w}_{N}(\Omega)$ appearing in the above formula has matrix
elements given explicitly by
\begin{equation}
\langle J,m_{2}|\hat{w}_{N}(\Omega)|J,m_{1}\rangle=\sqrt{4\pi}\sum_{j=0}^{2J}%
\frac{\sqrt{2j+1}}{2J+1}C_{Jm_{2},%
\,j(m_{1}-m_{2})}^{Jm_{1}}Y_{j,m_{1}-m_{2}}(\Omega),  \label{Eq:hatWelement}
\end{equation}
where $C_{Jm_{2},\,j(m_{1}-m_{2})}^{Jm_{1}}$ are Clebsch-Gordan coefficients
and $Y_{j,m}(\Omega)$ are the standard spherical harmonics functions. Let us
note that the highest degree of spherical harmonics occurring in the Wigner
function for the spin-$J$ system is $j=2J=N$. In contrast to the standard Wigner function defined in the position-momentum plane which is typically normalized to one \cite{Wigner1932}, we will follow the convention to normalize the spin Wigner function $W^N_{\hat{\varrho}}(\Omega)$ to $4\pi/(N+1)$ \cite{Varilly1989}. Crucially, the spin Wigner function defined above has the traciality property \cite{Stratonovich1956, Varilly1989}, that is, for every two operators $\hat{A},\,\hat{B}$ acting on a system with total angular momentum $J$ we have $\Tr[\hat{A}\hat{B}]=\frac{2J+1}{4\pi}\int_{\mathcal{S}_2}W_{\hat{A}}(\Omega)W_{\hat{B}}(\Omega)d\Omega$, where $W_{\hat{A}}(\Omega)$ and $W_{\hat{B}}(\Omega)$ are spin Wigner functions of operators $\hat{A}$ and $\hat{B}$ respectively, calculated according to Eq.~(\ref{eq:Wigner_general}).

\begin{figure}[th]
\includegraphics[width=1\columnwidth]{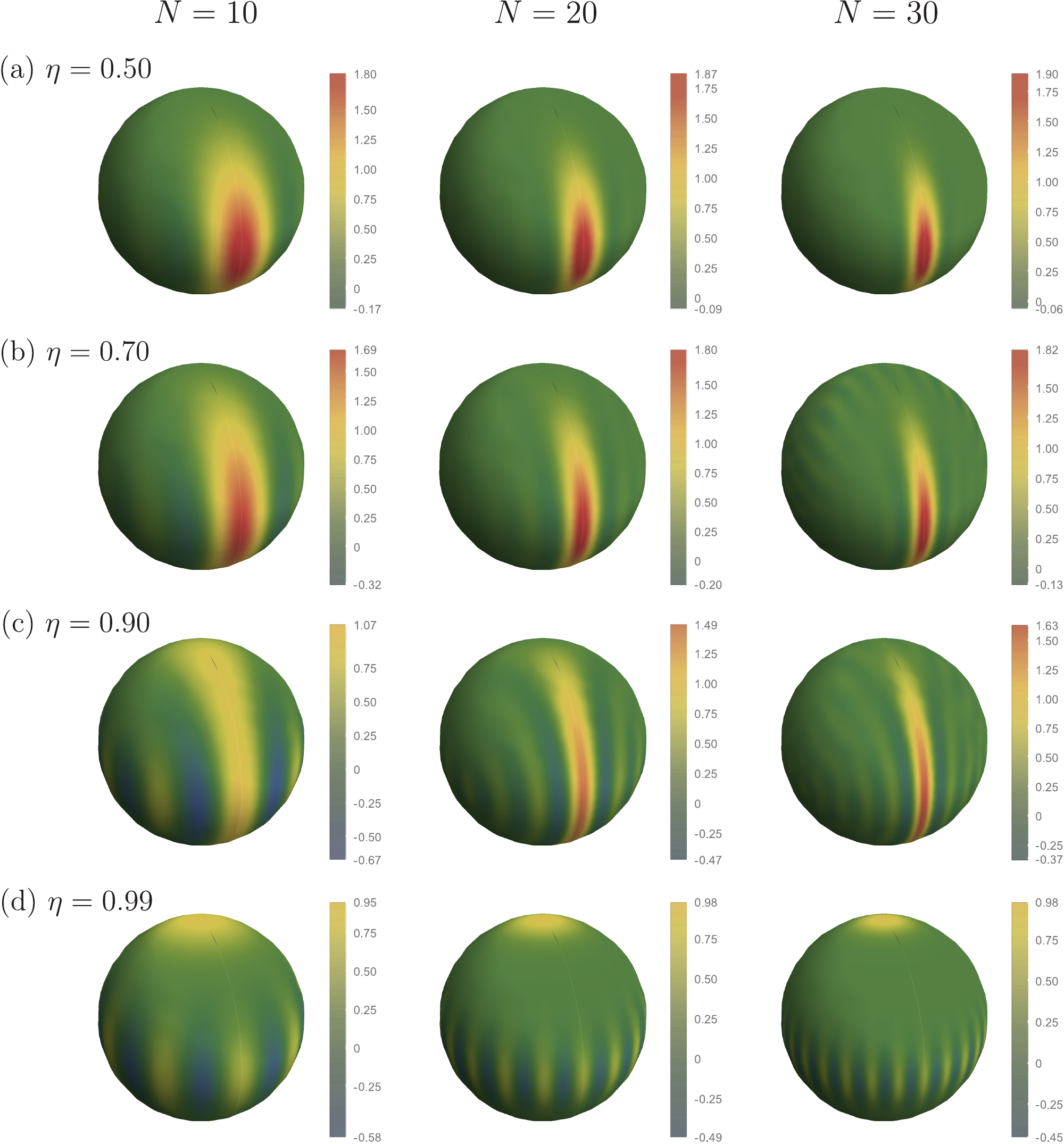}
\caption{The spin Wigner functions $W_{\hat{\protect\varrho}_{\mathrm{opt}%
}}^{N}(\protect\theta ,\protect\phi )$ for the optimal $N$-photon two-mode
input probe states for the three values of $N$ and the four values of photon
loss coefficient $\protect\eta $. Each function is accompanied by a legend,
which includes the minimum and maximum values that the function takes.}
\label{fig::optimalWigner}
\end{figure}

Fig.~\ref{fig::optimalWigner} depicts Wigner functions of the optimal input states
for the combinations of the photon number $N$ and the transmission $\eta$
marked in Fig.~\ref{Fig::loglog} with black dots. It should be noted that the numerical
optimization of the quantum Fisher information given in Eq.~(\ref{Eq:FQ=sumL}%
) leaves certain freedom regarding the choice of the phases of probability
amplitudes $c_m$ in the decomposition of the optimal states defined in Eq.~(%
\ref{Eq:psi}). In the numerical examples shown here, all the phases have
been set to zero in order to make the graphic representation most lucid.

Let us recall that the phase shift defined in Eq.~(\ref{eq:Uphi})
corresponds to rotation of the sphere $\mathcal{S}_2$ by an angle $\varphi$
about the vertical axis passing through the poles $\theta=0$ and $\pi$.
Intuitively, the sensitivity of the state to a phase shift, and therefore also precision of phase estimation, should be related
to how much the respective Wigner function changes with respect to such a
rotation. We will give a mathematical foundation to this intuition in \ref%
{Sec:Superfidelity} where we use the notion of superfidelity to derive a
general lower bound on the quantum Fisher information explicitly in terms of
the relevant Wigner functions.

It is seen in Fig.~\ref{fig::optimalWigner} that for the transmission
parameter $\eta=50\%$ the bulk of the quasiprobability is
concentrated on the equator of the sphere $\mathcal{S}_2$. The width of the
quasiprobability ``lump'' is noticeably reduced in the latitudinal direction,
along which the phase shift occurs, at the cost of expansion along the
longitude. This shape is reminiscent of the well known spin squeezed states \cite{Dowling1994}
that have been exploited in quantum enhanced atomic interferometry \cite{Kitagawa1993, Wineland1992, Ma2011}. With
decreasing losses, the form of the Wigner function qualitatively changes,
approaching for $\eta=99\%$ what can be identified as the representation of
the N00N state defined in Eq.~(\ref{Eq:N00Ndef}). The two quasiprobability
``lumps'' on the poles of the $\mathcal{S}_2$ sphere represent the component
states $\ket{J, \pm J}$ and the coherence between these components results
in a characteristic gear-like interference structure along the equator \cite{Schleich2001}.

\begin{figure}[tbp]
\includegraphics[width=\columnwidth]{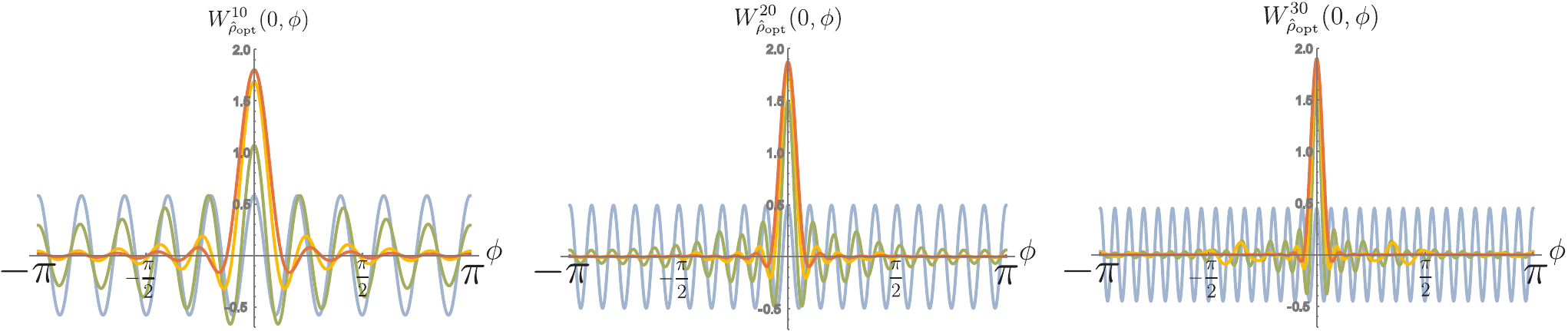}
\caption{The spin Wigner functions $W_{\hat{\protect\varrho}_{\mathrm{opt}%
}}^{N}(\protect\theta ,\protect\phi )$ for the optimal $N$-photon two-mode
input probe states with $\protect\theta =0$ for the three values of photon
number $N=10,\ 20\ \mathrm{and}\ 30$. The solid lines on each plot are
associated with particular values of photon loss coefficient $\protect\eta $%
. The color scheme for $\protect\eta $ is the same as in Figure \protect\ref{Fig::loglog}.}
\label{Fig::Wigneronequator}
\end{figure}

The transition between the squeezed-like regime and the Heisenberg-limited
regime can be visualized more clearly by plotting the cross-sections of the
Wigner functions along the equator, shown in Fig.~\ref{Fig::Wigneronequator}
using the same colour coding for the transmission parameter as in Fig.~\ref%
{Fig::loglog}. The gear-like structure for $\eta=99\%$ has the oscillation
period in $\phi$ equal to $2\pi/N$, which corresponds to the highest degree
of the spherical harmonics contributing to the Wigner function for a given $N
$. With increasing losses, sensitivity to the phase shift, and by this also precision of phase estimation, originates more
from the central peak of the quasiprobability distribution located at $\phi=0$.
This central peak can be considered as a superposition of spherical
harmonics with different oscillation frequencies along the equator of the $%
\mathcal{S}_2$ sphere.

\section{Phase space description of photon loss}
\label{sec:PhaseSpaceLoss}

In the preceding section we discussed qualitatively at the Wigner functions of
the optimal input states for lossy interferometry. The actual
precision of the interferometric measurement depends on how much the state
of light changes with respect to the phase shift after the photon loss. As
discussed in Sec.~\ref{Sec:LossTransformation}, the loss transformation maps
an $N$-photon state $\hat{\varrho}$ onto an ensemble of states given by $%
\hat{\varrho}^N_L = \Lambda^N_L(\hat{\varrho})$, with respective
probabilities $p_L^N$, where $L=0,\ldots, N$. The state $\hat{\varrho}^N_L$
is a two-mode state containing $N^{\prime }=N-L$ photons that remain in the
interferometer. Because there is no coherence left between subspaces with
different $N^{\prime }$ is is natural to consider the family of spin Wigner
functions $W^{N^{\prime }}_{\hat{\varrho}^N_{N-N^{\prime }}}(\theta,\phi)$
indexed with the remaining photon number $N^{\prime }$. This family can be
used to construct a lower bound on the quantum Fisher information. As we
show in \ref{Sec:Superfidelity}, the bound is given by the following
expression
\begin{equation}
F_Q[\hat{\varrho}_\varphi] \ge 2\sum_{N^{\prime }=0}^{N} p^N_{N-N^{\prime }}
\frac{N^{\prime }+1}{4\pi} \int_{\mathcal{S}_2} d\Omega \left( \frac{\partial%
}{\partial \phi} W^{N^{\prime }}_{\hat{\varrho}^N_{N-N^{\prime
}}}(\theta,\phi)\right)^2  \label{Eq:FQgesumlintS2}
\end{equation}
that involves the derivatives of the Wigner functions $W^{N^{\prime }}_{\hat{%
\varrho}^N_{N-N^{\prime }}}(\theta,\phi)$ with respect to the azimuthal
coordinate $\phi$. The integration over the sphere $\mathcal{S}_2$ takes the
standard form in the coordinates $(\theta,\phi)$:
\begin{equation}
\int_{\mathcal{S}_2} d\Omega = \int_{0}^{2 \pi} d\phi \int_{0}^{\pi}
\sin\theta\, d\theta.
\end{equation}
The above bound is a direct consequence of the traciality property of spin
Wigner functions.

\begin{figure}[tbp]
\includegraphics[width=\columnwidth]{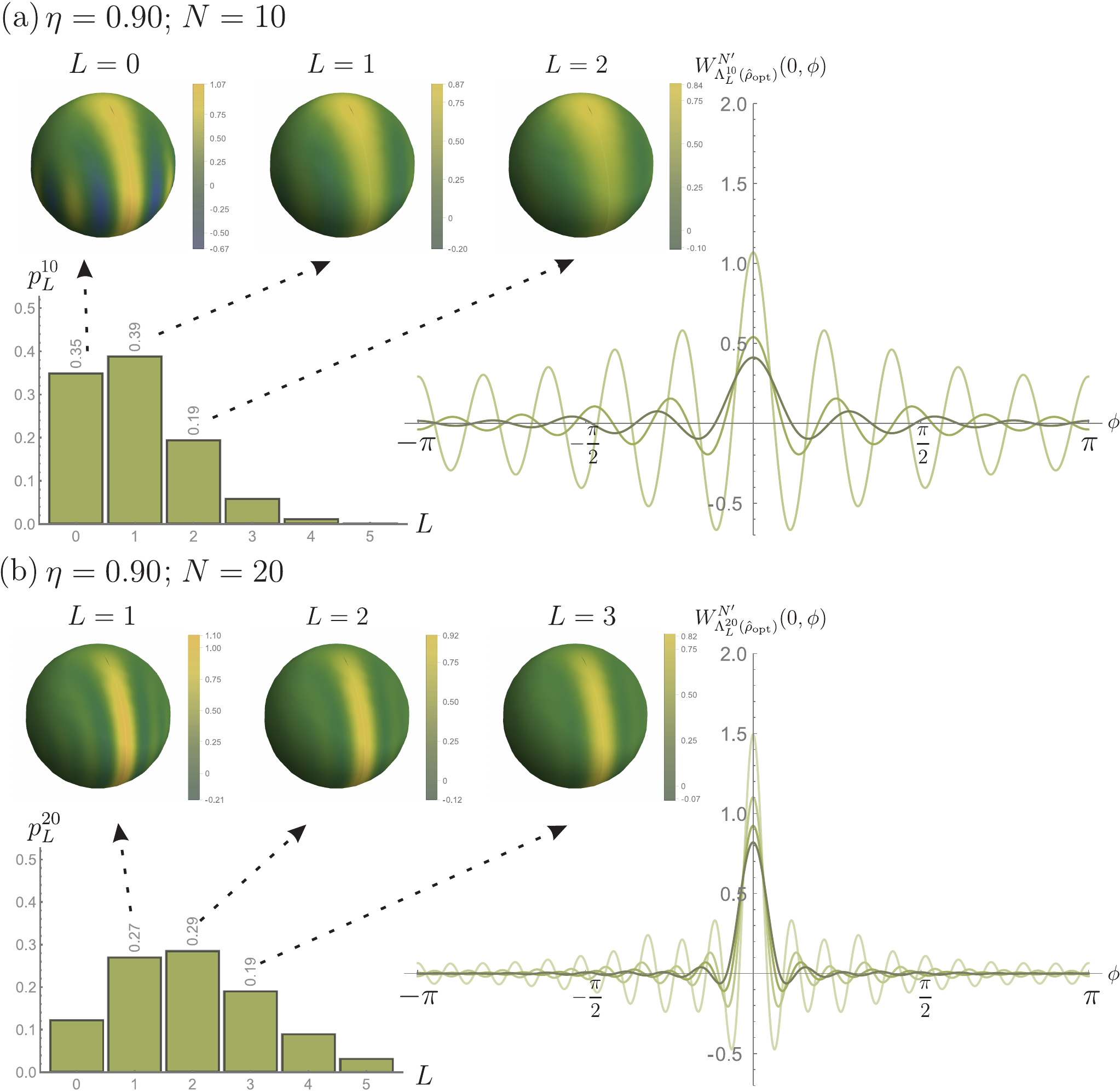}
\caption{The collected results for the angular Wigner functions $%
W_{\Lambda_{L}^{N}(\hat{\protect\varrho}_{\mathrm{opt}})}^{N^{\prime}}(%
\protect\theta,\protect\phi)$ associated with the loss of exactly $L$
photons for $\protect\eta=0.90$ and $N=10$ (a) and $N=20$ (b). The bar chart depicts the
binomial probability distribution $p_{L}^{N}$ as a function of the number
of lost photons $L$ with the three most probable values of $L=0,\ 1\
\mbox{and}\ 2$ for $N=10$ and $L=1,\,2,\,3$ for $N=20$ specified numerically. For each most probable $L$ we include the
spin Wigner functions and their representations on the equator with
different $L$s plotted in different shades of green. The shade is lightest for $L=0$ and becomes darker with L increased in unit steps.}
\label{Fig:Wignerafterloss}
\end{figure}

In order to gain further insight into the structure of optimal states, in
Fig.~\ref{Fig:Wignerafterloss} we plot the Wigner functions $W^{N^{\prime
}}_{\hat{\varrho}^N_L}(\Omega)$ of the conditional states for the
intermediate value of the loss parameter $\eta=90\%$, input photon numbers $%
N=10$ and $20$ and three most probable in each case numbers of lost photons $%
L$. It can be noticed that the loss transformation removes from the Wigner
function the spherical harmonics components with the highest degree.
Consequently, the gear-like equatorial structure characteristic for the N00N
state disappears even if a single photon is lost. In contrast, preparing a
superposition containing a range of spherical harmonics of varying degree
allows one to retain phase sensitivity in the presence of loss.

Instead of resorting to Eq.~(\ref{Eq:LossTransformation}) to describe the
conditional state after the photon loss, the loss transformation can be
represented as a convolution map on the $\mathcal{S}_2$ phase space. In
order to obtain an explicit expression, we will start from the Wigner
function for the state $\hat{\varrho}_L^N$ after loss
\begin{equation}
W_{\hat{\varrho}_{L}^{N}}^{N^{\prime }}(\Omega )=\Tr[\hat{w}_{N'}(\Omega)%
\Lambda_{L}^{N}(\hat{\varrho})],  \label{Eq:WLNNprime}
\end{equation}%
and use the inverse relation to represent the input state $\hat{\varrho}$ in
terms of its Wigner function,
\begin{equation}
\hat{\varrho}=\frac{N+1}{4\pi }\int_{\mathcal{S}_2} d\Omega^{\prime }\,W_{%
\hat{\varrho}}^{N}(\Omega^{\prime })\hat{w}_{N}(\Omega^{\prime }),
\label{Eq:varrhoinv}
\end{equation}
Inserting Eq.~(\ref{Eq:varrhoinv}) into Eq.~(\ref{Eq:WLNNprime}) yields the
convolution formula
\begin{equation}\label{eq:Wigner_convol}
W_{\hat{\varrho}_{L}^{N}}^{N^{\prime }}(\Omega )=\int d\Omega^{\prime }\,\mathcal{L}%
_{L}^{N}(\Omega ,\Omega ^{\prime })W_{\hat{\varrho}}^{N}(\Omega ^{\prime }),
\end{equation}%
with the integral kernel given by
\begin{equation}
\mathcal{L}_{L}^{N}(\Omega ,\Omega ^{\prime })=\frac{N+1}{4\pi }\Tr\bigl[%
\hat{w}_{N-L}(\Omega )\Lambda _{L}^{N}\bigl(\hat{w}_{N}(\Omega ^{\prime })%
\bigr)\bigr].  \label{Eq:L kernel}
\end{equation}%
Using the covariant form of the Wigner operator $\hat{w}_{N}(\Omega^{\prime
})=\hat{D}(\Omega^{\prime })\hat{w}_{N}(0)\hat{D}^{\dagger }(\Omega^{\prime
})$, where $\hat{D}(\Omega^{\prime })$ is the SU(2) displacement operator
and the argument $0$ denotes the north pole of the sphere $\mathcal{S}_2$,
one can simplify the expression for the integral kernel given in Eq.~(\ref%
{Eq:L kernel}) to the form
\begin{equation}
\mathcal{L}_{L}^{N}(\Omega ,\Omega ^{\prime })=\frac{N+1}{4\pi }\Tr\bigl[%
\hat{w}_{N-L}(\Omega ^{^{\prime }-1}\Omega )\Lambda _{L}^{N}\bigl(\hat{w}%
_{N}(0)\bigr)\bigr]=\mathcal{L}_{L}^{N}(\Omega ^{^{\prime }-1}\Omega ),
\end{equation}%
since the photon loss transformation commutes with SU(2) displacements \cite%
{Demkowicz2015}. Therefore we can keep only the first parameter of $\mathcal{%
L}_{L}^{N}$:
\begin{equation}
\mathcal{L}_{L}^{N}(\Omega )=\frac{N+1}{4\pi }\Tr\bigl[\hat{w}_{N-L}(\Omega
)\Lambda _{L}^{N}(\hat{w}_{N}(0))\bigr].  \label{kernelL}
\end{equation}
Because the operator $\hat{w}_{N}(0)$ is diagonal in the spin eigenbasis $%
\ket{J,m}$ and this property carries over to $\Lambda _{L}^{N}(\hat{w}%
_{N}(0))$, the integration kernel depends only on the polar angle $\theta$. Note that since $W^N_{\hat{\varrho}}(\Omega)$ and $W^{N^{\prime}}_{\hat{\varrho}^N_L}(\Omega)$ in Eq.~(\ref{eq:Wigner_convol}) are normalized to $4\pi/(N+1)$ and $4\pi/(N-L+1)$ respectively, the integration kernel is normalized to $\frac{N+1}{N-L+1}$.

In the context of the asymptotic precision limit for quantum-enhanced lossy
interferometry, it is interesting to consider the approximate form of the
integration kernel $\mathcal{L}_{L}^{N}(\Omega )$ in the regime when the
input photon number $N$, the lost photon number $L$, and the remaining
photon number $N^{\prime }=N-L$ are much greater than one. The derivation
presented in detail in \ref{Sec:AsymptoticsLossKernel} yields the following
asymptotic expression valid in this limit:
\begin{equation}\label{eq:Kernel_asympt}
\mathcal{L}_{L}^{N}(\Omega )\approx \frac{N+1}{2\pi }\left( \cos ^{2}\frac{%
\theta }{2}+\frac{L}{N}\sin ^{2}\frac{\theta }{2}\right) \mathcal{L}%
_{L}^{N}(\Omega )_{0},
\end{equation}%
where
\begin{equation}
\mathcal{L}_{L}^{N}(\Omega )_{0}\approx \frac{N+1}{L}\exp \left( -\frac{%
N(N-L)\sin ^{2}\theta }{2L}\right) .
\end{equation}%
For the most probable number of lost photons, when $L\approx (1-\eta )N$, it
is seen that the kernel $\mathcal{L}_{L}^{N}(\Omega )_{0}$ becomes approximately Gaussian around $\theta\approx 0$
with the exponential factor given by
\begin{equation}
\mathcal{L}_{L}^{N}(\Omega )_{0}\propto \exp \left( -\frac{\sin ^{2}\theta }{%
2\sigma ^{2}}\right),
\label{eq:kernel-asymp}
\end{equation}%
and the width equal to
\begin{equation}
\sigma =\sqrt{\frac{1-\eta }{\eta N}}
\end{equation}%
which is exactly the asymptotic precision bound for quantum interferometry
in the presence of two-arm photon loss given in Eq.~(\ref{eq:asymptotic_loss}%
).

\begin{figure}[tbp]
\centering \includegraphics[width=0.9\columnwidth]{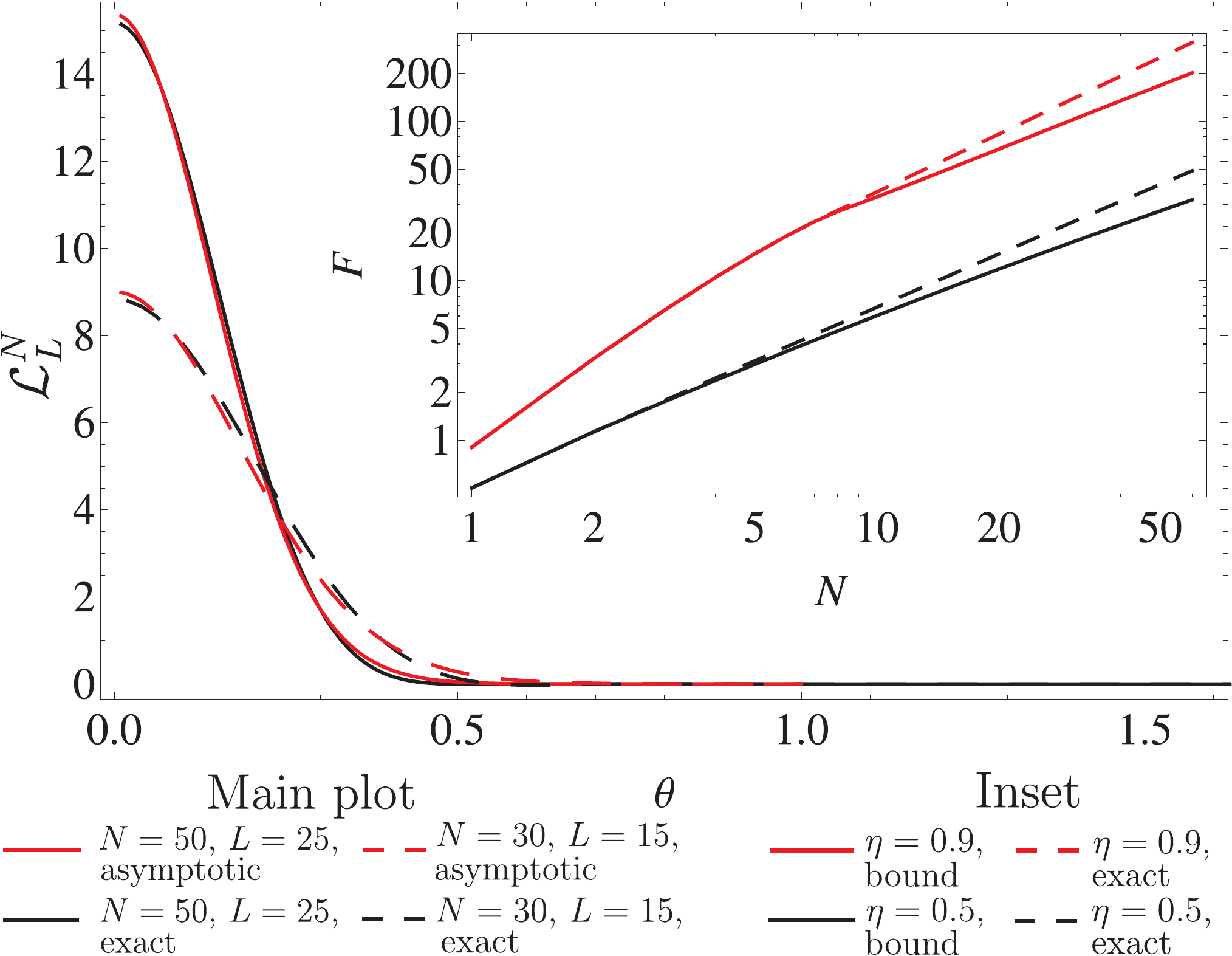}
\caption{The integration kernel $\mathcal{L}^N_L$ after normalization as a function of polar angle $\theta$ for $N=50,\,L=25$ (solid curves) and $N=30,\,L=15$ (dashed curves). Black curves represents the exact expression given in Eq.~(\ref{kernelL}) whereas red ones depict the asymptotic formula form Eq.~(\ref{eq:Kernel_asympt}). The inset shows the lower bound on the quantum Fisher information based Eq.~(\ref{Eq:FQgesumlintS2}) as a function of number of photons $N$ for two different loss coefficients $\eta=90\%$ (solid, red) and $\eta=50\%$ (solid, black) compared with the exact quantum Fisher information (analogously coloured dashed lines).
}
\label{fig::kernel}
\end{figure}

In Fig.~\ref{fig::kernel} we plot the integration kernel $\mathcal{L}^N_L$ as a function of polar angle for two different values of total number of photons $N=50$ and $N=30$ when the number of lost photons is equal to $L=N/2$. Such situation is meaningful for loss coefficient $\eta=50\%$. It can be seen that the exact formula Eq.~(\ref{kernelL}) converges to the asymptotic one Eq.~(\ref{eq:Kernel_asympt}), in particular the widths of respective curves become the same. Additionally, we plot the lower bound for the quantum Fisher information given by Eq.~(\ref{Eq:FQgesumlintS2}) as an inset in Fig.~\ref{fig::kernel} for two different loss coefficients $\eta=50\%$ and $\eta=90\%$.

The above result shows that in the asymptotic regime for the most likely
number of lost photons, $L \approx (1-\eta)N$, the oscillatory features of
the Wigner function whose characteristic size is of the order lower than $%
\sigma=\sqrt{(1-\eta)/\eta N}$ will be lost. In particular, for the N00N state the fringe period of
the gear-like structure on the equatorial ring is $2\pi/N$ and therefore it
will be completely washed out. The narrowest width of structures that can be
present in the Wigner function after the loss transformation is given by $%
\sigma$ and this scale defines the ultimate sensitivity limit of a lossy
interferometer. It is worth noting that the transmission-dependent factor $(1-\eta)/\eta$ is analogous to the one that characterizes "blurring" of the Wigner function of a single light mode sent through a beam splitter with transmission $\eta$ \cite{Leonhardt1993}. The additional factor $N$ stems from the convention of keeping the unit radius of the sphere $\mathcal{S}_2$ for any spin $N/2$.

\section{Conclusions}
\label{Sec:Conclusions}

In the lossless case, the optimal state for $N$-photon
two-mode quantum interferometry is the N00N state, which saturates the Heisenberg limit in the
phase-estimation protocol. The phase-space representation of the N00N state
in terms of the spin Wigner function shows a rich interference structure along the equator.
For substantial losses, the optimal states exhibit squeezing in the latitudinal direction
with the bulk of the quasiprobability located on the equator. Using the phase space picture,
we have shown how  optimal $N$-photon states
transit from the N00N states to spin-squeezed ones when the photon loss increases. We have found the integration
kernel which describes the transformation of the
phase-space distribution under photon loss. This transformation suppresses high-order spherical harmonics present
in the phase space distribution, which results the fragility of NOON
states with respect to photon losses. Furthermore, the width of the integration kernel
corresponds to the asymptotically attainable precision in the limit of large photon numbers.

\ack

We thank Rafa\l {} Demkowicz-Dobrza\'{n}ski for providing
us with a numerical procedure for calculating the coefficients of optimal $N$%
-photon two-mode input probe states. This work was supported by the European
Union Seventh Framework Programme (FP7/2007-2013) project PhoQuS@UW (Grant
agreement No. 316244), project CONACYT-CONICYT 208521 and project FONDECYT
1141178. This work is part of the project ``Quantum Optical
Communication Systems'' carried out within the TEAM
programme of the Foundation for Polish Science co-financed by the European Union under the European
Regional Development Fund.

\appendix

\section{Superfidelity bound}

\label{Sec:Superfidelity}

In order to assign a direct operational meaning to the Wigner function in
the context of quantum metrology we will derive a lower bound on the quantum
Fisher information containing solely quantities that can be expressed as
traces of quadratic formulas in terms of the density operator. Such traces
can be directly translated into phase space integrals of expressions that
are quadratic in the corresponding Wigner functions.

The bound will be based on a relation between the fidelity $\mathcal{F}(\hat{%
\varrho},\hat{\sigma}) = ||\sqrt{\hat{\varrho}}\sqrt{\hat{\sigma}}||^2_1$
between two arbitrary normalized states $\hat{\varrho}$ and $\hat{\sigma}$
and the so-called superfidelity, which is given by \cite{MiszczakQIP09}:
\begin{equation}
\mathcal{G} (\hat{\varrho} ,\hat{\sigma}) = \Tr(\hat{\varrho}\hat{\sigma}) +
\sqrt{1-\Tr(\hat{\varrho}^2)}\sqrt{1-\Tr(\hat{\sigma}^2)}.
\end{equation}
The superfidelity provides in general an upper estimate for the fidelity,
\begin{equation}
\mathcal{F} (\hat{\varrho} ,\hat{\sigma}) \le \mathcal{G} (\hat{\varrho} ,%
\hat{\sigma}).  \label{Eq:FleG}
\end{equation}
We will now apply the above inequality to a pair of infinitesimally close
states $\hat{\varrho}_\varphi$ and $\hat{\varrho}_{\varphi+\delta\varphi}$.
The left hand side can be related to quantum Fisher information through the
formula
\begin{equation}
F_Q[\hat{\varrho}_\varphi] (\delta\varphi)^2 = 4\left(1-\sqrt{\mathcal{F}(%
\hat{\varrho}_\varphi ,\hat{\varrho}_{\varphi+\delta\varphi})}\right).
\end{equation}
The above equation relates sensitivity of the state to small phase delays (measured by fidelity) and precision of phase estimation, given by quantum Fisher information.
On the right hand side of Eq.~(\ref{Eq:FleG}), we will expand the
superfidelity between the states $\hat{\varrho}_\varphi$ and $\hat{\varrho}%
_{\varphi+\delta\varphi}$ up to the second order in $\delta\varphi$, which
yields
\begin{eqnarray}
\mathcal{G}(\hat{\varrho}_\varphi ,\hat{\varrho}_{\varphi+\delta\varphi}) &
\approx & 1 - \frac{(\delta\varphi)^2}{2} \left\{ \Tr \left[ \left( \frac{%
\partial \hat{\varrho}_\varphi}{\partial \varphi} \right)^2 \right] + \frac{1%
}{1-\Tr(\hat{\varrho}_\varphi^2)} \left[\Tr \left( \hat{\varrho}_\varphi
\frac{\partial \hat{\varrho}_\varphi}{\partial\varphi} \right) \right]^2
\right\}  \nonumber \\
& &
\end{eqnarray}
Rearranging both sides of Eq.~(\ref{Eq:FleG}) implies that quantum Fisher
information is bounded from below by
\begin{equation}
F_Q[\hat{\varrho}_\varphi] \ge 2 \left\{ \Tr \left[ \left( \frac{\partial
\hat{\varrho}_\varphi}{\partial \varphi} \right)^2 \right] + \left( \frac{%
\partial}{\partial\varphi} \sqrt{1-\Tr(\hat{\varrho}_\varphi^2)} \right)^2
\right\}  \label{Eq:SuperfidelityBound}
\end{equation}
where we have converted the second term to a form that depends only on the
purity $\Tr(\hat{\varrho}_\varphi^2)$. Both terms are quadratic in density
matrices and can therefore be expressed as phase space integrals.

Let us now specialize the above general result to the phase measurement. For a
phase shift transformation the purity remains constant and therefore only
the first term in Eq.~(\ref{Eq:SuperfidelityBound}) is non-zero. Applying
the inequality to the Fisher information for the conditional state $\hat{%
\varrho}^{N}_{L}= \Lambda^{N}_{L}(\hat{\varrho}) $ after the loss of $L$
photons yields
\begin{equation}
F_Q[\hat{U}_\varphi \hat{\varrho}^{N}_{L} \hat{U}_\varphi^\dagger] \ge 2 \Tr %
\left[ \left( \frac{\partial}{\partial \varphi} \left(\hat{U}_\varphi \hat{%
\varrho}^{N}_{L} \hat{U}_\varphi^\dagger \right) \right)^2 \right].
\label{Eq:AppFQNLgeTr}
\end{equation}
Because the phase shift $\hat{U}_\varphi$ corresponds to the rotation of the
corresponding Wigner function about the vertical axis by $\varphi$, the
trace in the above formula can be written as
\begin{equation}
\Tr \left[ \left( \frac{\partial}{\partial \varphi} \left(\hat{U}_\varphi
\hat{\varrho}^{N}_{L} \hat{U}_\varphi^\dagger \right) \right)^2 \right] =
\frac{N-L+1}{4\pi} \int_{\mathcal{S}_2} d\Omega \left( \frac{\partial}{%
\partial \phi} W^{N-L}_{\hat{\varrho}^N_{L}}(\theta,\phi)\right)^2
\label{Eq:AppTr=int}
\end{equation}
where we have used the traciality property of spin Wigner functions.
Inserting Eqs.~(\ref{Eq:AppFQNLgeTr}) and (\ref{Eq:AppTr=int}) into Eq.~(\ref%
{Eq:FQ=sumL}) yields Eq.~(\ref{Eq:FQgesumlintS2}).

\section{Asymptotics of the loss transformation kernel}

\label{Sec:AsymptoticsLossKernel}

In order to obtain the approximate expression for the loss transformation
kernel
\begin{equation}
\mathcal{L}_{L}^{N}(\Omega )=\frac{N+1}{4\pi }\Tr\bigl[\hat{w}_{N-L}(\Omega
)\Lambda _{L}^{N}\bigl(\hat{w}_{N}(0)\bigr)\bigr]  \label{L}
\end{equation}%
in the limit of large initial number of photons $N\gg 1$ and losses
$L= (1-\eta) N$  so that $N^{\prime }=N-L\gg 1$,
we use the asymptotic expression for the angular momentum Wigner kernel in
the $N+1$ dimensional $SU(2)$ irreducible subspace \cite{KlimovJOSA00},
\begin{equation}
\hat{w}_{N}(\Omega )=\int d\omega f(\omega )e^{-i\omega \mathbf{\hat{J}}%
\cdot \mathbf{n}},
\end{equation}%
\begin{equation}
f(\omega )\simeq (-1)^{N/2}\left[ \delta (\omega -\pi )-\frac{i}{N/2}%
\partial _{\omega }\delta (\omega -\pi )\right] ,
\end{equation}%
where $\mathbf{n}=(\sin \theta \cos \phi ,\sin \theta \sin \phi ,\cos \theta
)$ and $\mathbf{\hat{J}}=(\hat{J}_{x},\hat{J}_{y},\hat{J}_{z})$ are angular
momentum operators. Performing a trivial integration we obtain
\begin{equation}
\hat{w}_{N}(\Omega )\approx (-1)^{N/2}\left( 1+\frac{2}{N}\mathbf{\hat{J}}%
\cdot \mathbf{n}\right) e^{-i\pi \mathbf{J}\cdot \mathbf{n}}.  \label{wn}
\end{equation}%
It follows from (\ref{Eq:lossinangularmomentumrep}) that the action of the
map $\Lambda _{L}^{N}$ on
\begin{equation}
\hat{w}_{N}(0)\approx (-1)^{N/2}\sum_{m=-N/2}^{N/2}\left( 1+\frac{2m}{N}%
\right) e^{-i\pi m}|N/2,m\rangle \langle N/2,m|,
\end{equation}%
has the form
\begin{eqnarray}
\Lambda _{L}^{N}(\hat{\omega}_{N}(0)) &=&(-1)^{N/2}{{{{\binom{N}{L}}}}}%
^{-1}\sum_{m=-N/2}^{N/2}\sum_{l=0}^{L}\left( 1+\frac{2m}{N}\right) e^{-i\pi
m}{{{{\binom{N/2+m}{l}}}}}  \nonumber \\
&\times &{{{{\binom{N/2-m}{L-l}}}}}|N^{\prime }/2,m+L/2-l\rangle \langle
N^{\prime }/2,m+L/2-l|,  \label{lc}
\end{eqnarray}%
being an operator acting in $N^{\prime }+1$ dimensional subspace.

Taking into account the form of $\hat{\omega}_{N}(\Omega )$ given in (\ref%
{wn}) we represent the kernel (\ref{L}) as a sum of four terms $\mathcal{L}%
_{L}^{N}(\Omega )_{k}$, $k=0,1,2,3$ of order $1$, $1/N$, $1/N^{\prime }$ and
$1/NN^{\prime }$ correspondingly
\begin{equation}
\mathcal{L}_{L}^{N}(\Omega )=\frac{N+1}{4\pi }\sum_{k=0}^{3}\mathcal{L}%
_{L}^{N}(\Omega )_{k}.  \label{Ltot}
\end{equation}%
After a long but straightforward algebra one obtains
\begin{eqnarray}
\mathcal{L}_{L}^{N}(\Omega )_{k} &=&(N+1)(N^{\prime })!\sum_{m=-N^{\prime
}/2}^{N^{\prime }/2}(-1)^{N^{\prime }+2m}d_{mm}^{N^{\prime }/2}(2\theta
)\sum_{l=0}^{L}(-1)^{l}{{{{\binom{L}{l}}}}}  \nonumber \\
&\times &\frac{B(N^{\prime }/2+m+l+1,N^{\prime }/2+L-m-l+1)}{(N^{\prime
}/2+m)!(N^{\prime }/2-m)!}g_{k}(m,L,l),
\end{eqnarray}%
where $B(x,y)=\Gamma (x)\Gamma (y)/\Gamma (x+y)$ is the $B$-function, $%
d_{mm}^{N}(\theta )$ is the Wigner $d$-function from $N+1$ dimensional
representation of the $SU(2)$ group,
\begin{equation}
\langle N,m|e^{-i\theta J_{y}}|N,n\rangle =d_{mn}^{N}(\theta ),
\end{equation}%
and
\begin{eqnarray*}
g_{0}(m,L,l) &=&1,\ \ g_{1}(m,L,l)=\frac{2}{N}(m+l-L/2), \\
g_{2}(m,L,l) &=&\frac{2}{N^{\prime }}l\sec \theta ,\ \ g_{3}(m,L,l)=\frac{4}{%
NN^{\prime }}(m+l-L/2)l\sec \theta .
\end{eqnarray*}%
\textit{Asymptotic expansion of $\mathcal{L}_{L}^{N}(\Omega )_{0}$} Using
integral representations for the $B$-function and the Wigner $d$-function
\cite{Varshalovich88} we rewrite $\mathcal{L}_{L}^{N}(\Omega )_{0}$ as
\begin{eqnarray}
\mathcal{L}_{L}^{N}(\Omega )_{0} &=&\frac{N+1}{4\pi }\int_{-1}^{1}dt%
\int_{0}^{2\pi }d\varphi \,t^{L}\left[ \cos \theta +\sin \theta \left( i\cos
\varphi +t\sin \varphi \right) \right] ^{N-L},  \nonumber \\
&&
\end{eqnarray}%
which due to its parity property is non-zero only for $L=2K$ and can be
reduced to the following form
\begin{eqnarray}
\mathcal{L}_{2K}^{N}(\Omega )_{0} &=&\frac{N+1}{2\sqrt{\pi }}\frac{\Gamma
(K+1/2)}{\Gamma (K+1)}\int_{-1}^{1}dx\,(1-x^{2})^{K}\left( \cos \theta
+ix\sin \theta \right) ^{N-2K}  \nonumber \\
&=&\frac{(N+1)\Gamma (K+1/2)2^{K-3/2}}{\sin ^{K+1/2}\theta }%
P_{N-K+1/2}^{-K-1/2}(\cos \theta ),  \label{P2}
\end{eqnarray}%
where $P_{\mu }^{\nu }(\cos \theta )$ is the associated Legendre function.
At $K=0$\ (zero losses) one recovers $\mathcal{L}_{0}^{N}(\Omega )_{0}$\ as
a character of $N+1$\ dimensional representation of the $SU(2)$\ group,
\[
\mathcal{L}_{0}^{N}(\Omega )_{0}=\frac{\sin (N+1)\theta }{\sin \theta }.
\]

In the asymptotic limit $N-2K\gg 1$, the kernel $L_{2K}^{N}(\Omega )_{0}$\
oscillates for small values of $K$, and tends to a Gaussian function when $%
K\gg 1$,
\[
\mathcal{L}_{2K}^{N}(\Omega )_{0}\approx \frac{N+1}{2K}\exp \left[ -\frac{%
N(N-2K)\sin ^{2}\theta }{4K}\right] .
\]

\noindent The remaining terms of $\mathcal{L}_{L}^{N}(\Omega )$ can be
evaluated in a very similar way, in particular,
\begin{equation}
\mathcal{L}_{L}^{N}(\Omega )_{1}=\frac{N-L}{N}\cos \theta \mathcal{L}%
_{L+1}^{N-1}(\Omega )_{0}+\frac{L}{N}\mathcal{L}_{L-1}^{N}(\Omega )_{0}+%
\mathcal{O}(\sin \theta ),  \label{L1}
\end{equation}%
\begin{equation}
\mathcal{L}_{L}^{N}(\Omega )_{2}=\mathcal{L}_{L+1}^{N-1}(\Omega )_{0}+%
\mathcal{O}(\sin \theta ),  \label{L2}
\end{equation}%
where $L=2K+1$, and
\begin{eqnarray}
\mathcal{L}_{L}^{N}(\Omega )_{3} &=&\frac{N-L}{N}\cos \theta \mathcal{L}%
_{L+2}^{N-2}(\Omega )_{0}+\frac{1}{N}\cos \theta \left( \mathcal{L}%
_{L}^{N-2}(\Omega )_{0}-\mathcal{L}_{L+2}^{N-2}(\Omega )_{0}\right)
\nonumber \\
&+&\frac{L}{N}\mathcal{L}_{L}^{N-1}(\Omega )_{0},  \label{L3}
\end{eqnarray}%
where $L=2K$. Expressions given in (\ref{L1})-(\ref{L3}) simplify for large
losses, $L\gg 1$. Consequently, $\mathcal{L}_{L}^{N}(\Omega )$ acquires the
form
\begin{equation}
\mathcal{L}_{L}^{N}(\Omega )=\frac{N+1}{4\pi }\left[ 1+\frac{N-2K}{N}\cos
\theta +\frac{2K}{N}+\mathcal{O}(1/N)\right] \mathcal{L}_{2K}^{N}(\Omega
)_{0},  \label{L final}
\end{equation}%
both for even and odd number of lost photons $L$.

\bibliographystyle{iopart-num}
\bibliography{biblio_phase_space}

\end{document}